\documentclass[final,11pt]{article}

\usepackage{multirow}
\usepackage[nottoc]{tocbibind}
\usepackage{geometry}
\usepackage{subcaption}
\usepackage[affil-it, auth-sc]{authblk}
\usepackage[english]{babel}
\usepackage[utf8]{inputenc}
\usepackage{graphicx}
\usepackage{amssymb}
\usepackage{amsthm}
\usepackage{amsmath}
\usepackage{amsfonts}
\usepackage{verbatim}
\usepackage{lineno}
\usepackage{todonotes}
\presetkeys{todonotes}{inline, color=blue!30, size=\small}{}
\usepackage{color, soul}
\definecolor{darkred}{rgb}{0.9, 0.0, 0.0}
\definecolor{darkgreen}{rgb}{0.0, 0.5, 0.0}

\usepackage[colorlinks,bookmarks,linkcolor=darkred, citecolor=darkgreen]{hyperref}
\usepackage{eso-pic}
\usepackage[sort&compress,numbers]{natbib}
\usepackage{doi}
\usepackage{paralist,epsfig}
\usepackage{relsize}
\usepackage{nicefrac,esint}
\usepackage{float}
\usepackage{cleveref}

\begin{document}

\AddToShipoutPictureFG*{
    \AtPageUpperLeft{\put(-60,-60){\makebox[\paperwidth][r]{LA-UR-23-30368}}}  
    }

\title{Medium-induced photon bremsstrahlung in neutrino-nucleus, antineutrino-nucleus, and electron-nucleus scattering from multiple QED interactions}

\author{Oleksandr Tomalak\thanks{tomalak@lanl.gov}}
\author{Ivan Vitev\thanks{ivitev@lanl.gov}}
\affil{Theoretical Division, Los Alamos National Laboratory, Los Alamos, NM 87545, USA}

\date{\today}

\maketitle
Interactions of charged leptons with nuclei and the naive tree-level kinematics of these processes are affected by radiation of photons induced by the QED nuclear medium. We evaluate cross-section modifications at leading orders of the number of correlated interactions inside the nucleus, known as the opacity expansion. We derive results for soft and collinear types of the bremsstrahlung at the first three orders in opacity and generalize them to higher orders. We present the leading in opacity energy spectra of soft and collinear photons and radiative energy loss inside the nucleus for experiments with lepton kinematics in the GeV energy range. At leading power of the Glauber soft-collinear effective field theory, the soft radiation is further resummed to all orders both in opacity and in the electromagnetic coupling constant. We find that the soft and collinear medium-induced radiation is vacuumlike, and additional corrections are power suppressed. Despite the negligible modification to the induced photon spectra, the nuclear medium-induced radiation sizably affects the broadening of charged leptons in the direction orthogonal to their propagation.

\newpage
\tableofcontents
\newpage

\section{Introduction}

Quantum electrodynamics (QED) radiative corrections are an integral part of the scattering cross sections that enter precise measurements of the nonperturbative quantities  describing the internal quantum chromodynamics (QCD) structure of hadrons and nuclei. Only with a proper unfolding of QED effects can precision experiments assess the process-independent QCD quantities. The development of techniques to evaluate QED radiative corrections~\cite{Yennie:1961ad,Mo:1968cg,Maximon:2000hm,Vanderhaeghen:2000ws,Gramolin:2014pva,Afanasev:2023gev}, fortunately,  accompanied the progress in electron scattering experiments~\cite{Dombey:1969wk,Akhiezer:1974em,A1:2010nsl,A1:2013fsc,Xiong:2019umf,Punjabi:2015bba,Bosted:1989hy,Perdrisat:2006hj,JeffersonLabHallA:2022cit,JeffersonLabHallA:2022ljj}. In the electroweak sector, the ambitious goals of the next generation of neutrino oscillation experiments~\cite{ESSnuSB:2013dql,DUNE:2016ymp,DUNE:2020ypp,Hyper-Kamiokande:2016dsw} motivated research on radiative corrections to the scattering of (anti)neutrino beams~\cite{DeRujula:1979grv,Day:2012gb,Tomalak:2021hec,Tomalak:2022xup}. Moreover, future measurements of the internal structure of nucleons and nuclei at the Electron-Ion Collider~\cite{Accardi:2012qut,AbdulKhalek:2021gbh} need proper treatment of QED radiative corrections that become sizable due to the enhancement by large logarithms of kinematic quantities~\cite{Liu:2020rvc}.

In performing experiments with nuclei, QED interactions of initial- and final-state particles with the nuclear medium in (anti)neutrino-, electron-, and muon-nucleus scattering might modify experimental observables in a very distinct way. Undoubtedly, these process-dependent corrections should be investigated in detail and applied in the analyses of measurements if the nuclear medium-induced effects are at the level of experimental uncertainty or above.

In our previous work~\cite{Tomalak:2022kjd}, we performed the first evaluation of cross-section modifications for the elastic scattering off nucleons\footnote{By elastic scattering on nucleons, we denote the process with initial and final states consisting of the lepton and the nucleon and allow for change of the kinematics.} inside a large nucleus due to the exchange of photons between charged particles and the nuclear medium in (anti)neutrino-, electron-, and muon-induced reactions. We found permille-level effects in (anti)neutrino-nucleus scattering and up to percent-level corrections in electron-nucleus scattering in forward kinematics~\cite{Tomalak:2022kjd}. We employed two approaches: the first one accounts only for the leading power of the soft-collinear effective field theory with Glauber-photon exchanges between the charged lepton and the QED nuclear medium ($\mathrm{SCET_G}$), while the second one is complete and does not rely on the approximations of $\mathrm{SCET_G}$. Subsequently, we included the multiple rescattering of charged leptons in a large nucleus within the effective field theory approach~\cite{Tomalak:2023kwl} by evaluating a well-defined series in the mean number of scatterings inside the nucleus, known as the opacity expansion, and found sizable QED nuclear medium-induced redistribution of charged particle tracks in the direction transverse to their propagation. The size of the corresponding corrections to the (anti)neutrino-nucleus and electron-nucleus cross sections can reach percent level.

In this work, we take the next step in the studies of QED nuclear medium effects and evaluate soft and collinear bremsstrahlung induced by the interaction of leptons with the nuclear medium mediated by Glauber photons. We derive radiative cross sections at the first three orders in the opacity expansion and generalize them to an arbitrary order. We show that soft radiation can be exactly resummed up to all orders in opacity at leading power of the soft-collinear effective field theory. We find that the resulting photon energy spectrum coincides with the spectrum in vacuum, with negligibly small power-suppressed corrections from the QED nuclear medium. We reproduce the expressions for soft radiation in the appropriate limit of our calculation for the collinear bremsstrahlung, and find that the relative contribution of nuclear medium-induced radiation to the bremsstrahlung in vacuum is approximately the same for soft and collinear photons at each order of the opacity expansion. At leading order of the opacity expansion, we provide numerical estimates for the nuclear medium-induced spectral distortion of the collinear radiation. Based on the derived nuclear medium-induced bremsstrahlung, we estimate the radiative energy loss of charged leptons inside the nucleus for incoming beams in the GeV energy range. Even though we apply similar theoretical tools for the derivation of medium-induced radiation in QED, our results differ both qualitatively and quantitatively from the QCD case.

The rest of our paper is organized as follows. In Section~\ref{sec:soft_section}, we derive the nuclear medium-induced emission of soft photons. First, we present the scattering amplitudes and unpolarized cross sections for the bremsstrahlung in Subsection~\ref{sec:soft_opacity}. Subsequently, we resum all orders in the opacity expansion for the soft radiation in Subsection~\ref{sec:soft_all_orders}. In this Subsection, we also present the soft functions that modify the broadening distributions in (anti)neutrino- and electron-nucleus scattering and present numerical estimates for beams of GeV energy. Section~\ref{sec:collinear_section} is dedicated to collinear bremsstrahlung. We present the opacity expansion for the collinear radiation in Subsection~\ref{sec:collinear_opacity}. We discuss the soft-photon, hard-photon, and phase-independent limits and generalize the opacity resummation to the case of collinear radiation in Subsection~\ref{sec:collinear_all_orders}. In Subsection~\ref{sec:collinear_first_opacity_order_numerics}, we present numerical results for the QED nuclear medium-induced radiation at GeV energies of the charged particle at the first order of the opacity expansion for several nuclei of interest to the neutrino and electron scattering communities. In Subsection~\ref{sec:collinear_QCD}, we discuss similarities and differences to the medium-induced radiation of gluons in QCD. We define and evaluate the radiative energy loss of the charged lepton in vacuum and in the nuclear medium in Section~\ref{sec:energy_loss}.  Section~\ref{sec:conclusions} contains the conclusions of our work. In Appendix~\ref{app:derivation}, we describe important details for the diagrammatic derivation of medium-induced cross-section corrections starting with expressions for the scattering amplitude. Appendixes~\ref{app:limit_large_x} and~\ref{app:limit_intermediate_x} describe various limits of the medium-induced radiation of collinear photons.

\section{Radiation of soft photons} \label{sec:soft_section}

In this Section, we evaluate cross sections for the radiation of soft photons. We derive analytic expressions for the first three orders of the opacity expansion, present the generalization to higher orders, and resum all orders of the opacity and electromagnetic coupling constant, but at the leading power of $\mathrm{SCET}_\mathrm{G}$.

\subsection{Opacity expansion of the soft-photon radiation} \label{sec:soft_opacity}

Let us consider the medium-induced radiation of one real soft photon and evaluate matrix elements for diagrams that contribute to the first orders of the opacity expansion.

After the hard scattering, the charged lepton propagates in the Coulomb field of electric charges inside the nucleus with a potential that in momentum space reads $v \left( {q}_\perp \right) = \frac{4 \pi \alpha}{q^2_{\perp} + \zeta^2}$, where $\zeta \approx \frac{m_e Z^{1/3}}{192}$~\cite{Jackson:1998nia} introduces the regularization at the atomic screening scale, $Z$ is the nuclear charge, $m_e$ is the electron mass, $\alpha$ is the electromagnetic coupling constant, and we denote $a_\perp = |\vec{a}_{\perp} |$ for an arbitrary vector $a$. The radiation of one soft photon from the final-state charged lepton at the space-time point $x^0$ multiplies the tree-level amplitude in vacuum $\mathrm{T}^\mathrm{LO}$ as
\begin{align}
\mathrm{T}_{1 \gamma}^\mathrm{LO} = e  \frac{ p^\prime \cdot \varepsilon^\star}{p^\prime \cdot k_\gamma} \mathrm{T}^\mathrm{LO} e^{i k_\gamma \cdot x^0}, \label{eq:soft_radiation}
\end{align}
where $\mathrm{T}_{1 \gamma}^\mathrm{LO}$ is now the leading-order amplitude with  radiation, $e = \sqrt{4 \pi \alpha}$ is the electric charge, $p^\prime$ is the momentum of the charged lepton, and $k_\gamma$ and $\varepsilon$ are the momentum and the polarization vector of the radiated photon, respectively. The amplitude in Eq.~(\ref{eq:soft_radiation}) is not gauge invariant. To obtain gauge-invariant radiation, we add an extra contribution from the radiation of heavy spectator with a four-velocity vector $v^\mu$,
\begin{align}
\mathrm{T}_{1 \gamma}^\mathrm{LO} = e \left( \frac{ p^\prime \cdot \varepsilon^\star}{p^\prime \cdot k_\gamma} - \frac{ v \cdot \varepsilon^\star}{v \cdot k_\gamma} \right) \mathrm{T}^\mathrm{LO} e^{i k_\gamma \cdot x^0}. \label{eq:soft_radiation_momenta_GI}
\end{align}

\begin{figure}[ht]
    \vspace{1.0cm}
    \centering
    \includegraphics[height=0.15\textwidth]{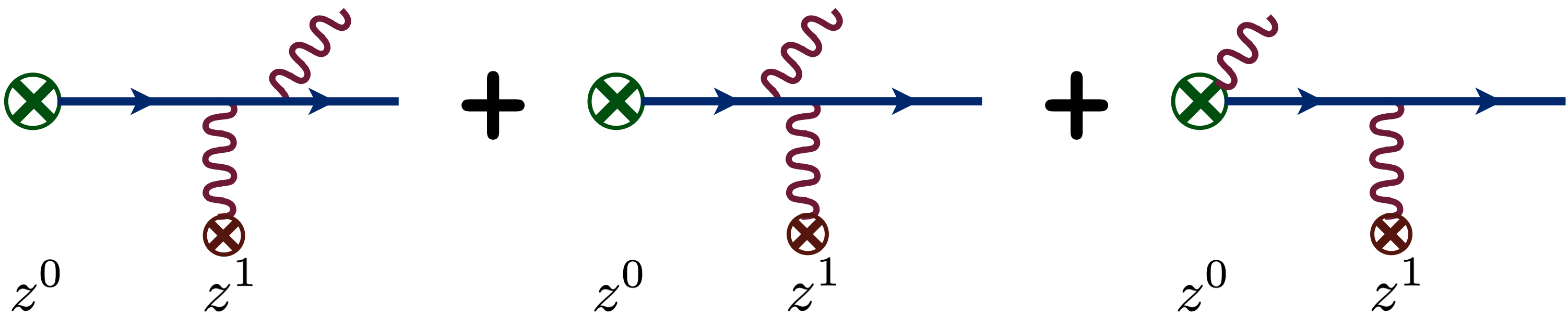}
    \caption{Diagrams for the one-Glauber exchange amplitude with one radiated photon in the charged-current (anti)neutrino-nucleus scattering.}\label{fig:one_photon_one_Glauber}
\end{figure}

An exchange of one Glauber photon with the nuclear medium at the spatial position $\left( \vec{x}^1_\perp, z^1 \right)$ gives an extra contribution $\mathrm{T}_{1 \gamma}^\mathrm{1G}$ to the matrix element $\mathrm{T}_{1 \gamma}^\mathrm{LO}$ of the hard process at the space-time point $x^0$
\begin{align}
\mathrm{T}_{1 \gamma}^\mathrm{1G} &= - i e  \left( \frac{ p^\prime \cdot \varepsilon^\star}{p^\prime \cdot k_\gamma} - \frac{ v \cdot \varepsilon^\star}{v \cdot k_\gamma} \right) \sum_{z^1 > z^0} \int \frac{\mathrm{d}^2 \vec{q}_\perp}{\left( 2 \pi \right)^2}  e^{ - i \vec{q}_{\perp} \cdot  \vec{x}^1_{\perp}} {v} \left( {q}_\perp \right)  \Gamma^{(1)} \left( \vec{q}_{\perp}, z^1 \right) \mathrm{T}^\mathrm{LO} \left( p^\prime + k_\gamma - q \right) e^{ i \left( p^\prime + k_\gamma   \right)  \cdot x^0 }, \\
\Gamma^{(1)} &= e^{i \Omega \left( p^\prime + k_\gamma , \vec{q}_{\perp} \right) \left( z^1 - z^0 \right)}   -  \frac{2 p^\prime \cdot k_\gamma}{\left( p^\prime+k_\gamma\right)_+} \frac{ e^{i \Omega \left( p^\prime , \vec{q}_{\perp} \right) \left( z^1 - z^0 \right)} - e^{i \Omega \left( p^\prime + k_\gamma , \vec{q}_{\perp} \right) \left( z^1 - z^0 \right)}}{\Omega \left( p^\prime , \vec{q}_{\perp} \right) - \Omega \left( p^\prime + k_\gamma , \vec{q}_{\perp} \right) }, \end{align}
with the phase factor $\Omega \left(p, \vec{q}_{\perp} \right)$,
\begin{align}
&\Omega \left(p, \vec{q}_{\perp} \right) = p^- -  \frac{ \left( \vec{p}_\perp - \vec{q}_\perp \right)^2}{p^+}. \label{eq:amplitude_soft_first_order}
\end{align}
We illustrate the corresponding diagrams in Fig.~\ref{fig:one_photon_one_Glauber}. We select the $z^1,~z^0$ coordinate axes and the third component of $\left( \vec{x}^1_\perp, z^1 \right),~\left( \vec{x}^0_\perp, z^0 \right)$ along the charged lepton momentum direction, respectively, and use standard light-cone coordinates.

An exchange of two Glauber photons with the nuclear medium at the spatial positions $\left( \vec{x}^1_\perp, z^1 \right)$ and $\left( \vec{x}^2_\perp, z^2 \right)$ gives an extra contribution $\mathrm{T}_{1 \gamma}^\mathrm{2G}$ to the matrix element $\mathrm{T}_{1 \gamma}^\mathrm{LO}$ of the hard process at the space-time point $x^0$,
\begin{align}
\mathrm{T}_{1 \gamma}^\mathrm{2G} &= -  e \left( \frac{ p^\prime \cdot \varepsilon^\star}{p^\prime \cdot k_\gamma} - \frac{ v \cdot \varepsilon^\star}{v \cdot k_\gamma} \right) \sum_{z^2 > z^1 > z^0} \int \frac{\mathrm{d}^2 \vec{q}_{1,\perp}}{\left( 2 \pi \right)^2} \frac{\mathrm{d}^2 \vec{q}_{2,\perp}}{\left( 2 \pi \right)^2} e^{ - i \left(\vec{q}_{1,\perp} \cdot \vec{x}^1_{\perp} + \vec{q}_{2,\perp} \cdot  \vec{x}^2_{\perp}\right)}\nonumber \\
& \times  {v} \left( q_{1,\perp} \right) {v} \left( q_{2,\perp} \right) \Gamma^{(2)} \left( \vec{q}_{1,\perp}, \vec{q}_{2,\perp}, z^1, z^2 \right) \mathrm{T}^\mathrm{LO} \left( p^\prime + k_\gamma - q_1 - q_2 \right) e^{ i \left( p^\prime + k_\gamma \right)  \cdot x^0 }, \\
\Gamma^{(2)} &=   e^{i \Omega \left( p^\prime + k_\gamma, \vec{q}_{1,\perp} + \vec{q}_{2,\perp} \right) \left( z^1 - z^0 \right)} e^{i \Omega \left( p^\prime + k_\gamma, \vec{q}_{2,\perp} \right) \left(z^2 - z^1 \right) } \nonumber \\
& -  \frac{2 p^\prime \cdot k_\gamma}{\left( p^\prime+k_\gamma\right)_+} e^{i \Omega \left( p^\prime + k_\gamma, \vec{q}_{1,\perp} + \vec{q}_{2,\perp} \right)  \left( z^1 - z^0 \right)} \frac{ e^{i \Omega \left( p^\prime, \vec{q}_{2,\perp} \right) \left( z^2 - z^1 \right) } - e^{i \Omega \left( p^\prime + k_\gamma, \vec{q}_{2,\perp} \right) \left(z^2 - z^1 \right) } }{ \Omega \left( p^\prime, \vec{q}_{2,\perp} \right) -  \Omega \left( p^\prime + k_\gamma, \vec{q}_{2,\perp} \right)}\nonumber \\
&-  \frac{2 p^\prime \cdot k_\gamma}{\left( p^\prime+k_\gamma\right)_+} \frac{ e^{i \Omega \left( p^\prime, \vec{q}_{1,\perp} + \vec{q}_{2,\perp} \right)  \left( z^1 - z^0 \right)} - e^{i \Omega \left( p^\prime + k_\gamma, \vec{q}_{1,\perp} + \vec{q}_{2,\perp} \right)  \left( z^1 - z^0 \right) } }{ \Omega \left( p^\prime, \vec{q}_{1,\perp} + \vec{q}_{2,\perp} \right) -  \Omega \left( p^\prime + k_\gamma, \vec{q}_{1,\perp} + \vec{q}_{2,\perp} \right)} e^{i \Omega \left( p^\prime, \vec{q}_{2,\perp} \right) \left(z^2 - z^1 \right) }.
\end{align}

After deriving the third order of the opacity expansion, we generalize the amplitude expression to higher orders and present the $n$th-order result below. An exchange of $n$ Glauber photons with the nuclear medium at the spatial positions $\left( \vec{x}^1_\perp, z^1 \right)$, $...,$ $~\left( \vec{x}^n_\perp, z^n \right)$ gives an extra contribution $\mathrm{T}_{1 \gamma}^\mathrm{nG}$ to the matrix element $\mathrm{T}_{1 \gamma}^\mathrm{LO}$ of the hard process at the space-time point $x^0$,
\begin{align}
\mathrm{T}_{1 \gamma}^\mathrm{nG} &= \left(- i\right)^n e \left( \frac{ p^\prime \cdot \varepsilon^\star}{p^\prime \cdot k_\gamma} - \frac{ v \cdot \varepsilon^\star}{v \cdot k_\gamma} \right)  \sum_{z^n > ... > z^1 > z^0} \int \frac{\mathrm{d}^2 \vec{q}_{1,\perp}}{\left( 2 \pi \right)^2} ... \frac{\mathrm{d}^2 \vec{q}_{n,\perp}}{\left( 2 \pi \right)^2} e^{ - i \left(\vec{q}_{1,\perp} \cdot  \vec{x}^1_{\perp} + ... + \vec{q}_{n,\perp} \cdot  \vec{x}^n_{\perp}\right)} {v} \left( q_{1,\perp} \right) ... {v} \left( q_{n,\perp} \right)  \nonumber \\
&  \times \Gamma^{(n)} \left( \vec{q}_{1,\perp},..., \vec{q}_{n,\perp}, z^1, ..., z^n \right) \mathrm{T}^\mathrm{LO} \left( p^\prime + k_\gamma - q_1 -... - q_n \right) e^{ i \left( p^\prime + k_\gamma   \right)  \cdot x^0 }, \\
\Gamma^{(n)} &=   e^{i \Omega \left( p^\prime + k_\gamma, \vec{q}_{1,\perp} + ... + \vec{q}_{n,\perp} \right) \left( z^1 - z^0 \right)} ... e^{i \Omega \left( p^\prime + k_\gamma, \vec{q}_{n-1,\perp} + \vec{q}_{n,\perp} \right) \left(z^{n-1} - z^{n-2} \right) } e^{i \Omega \left( p^\prime + k_\gamma,  \vec{q}_{n,\perp} \right) \left(z^n - z^{n-1} \right) }  \nonumber \\
&- \frac{2 p^\prime \cdot k_\gamma}{\left( p^\prime + k_\gamma \right)^+} \sum \limits_{i = 1}^{n} e^{i \Omega \left( p^\prime + k_\gamma, \vec{q}_{1,\perp} + ... + \vec{q}_{n,\perp} \right) \left( z^1 - z^0 \right)} ... e^{i \Omega \left( p^\prime + k_\gamma, \vec{q}_{i-1,\perp} + ...+ \vec{q}_{n,\perp} \right) \left(z^{i-1} - z^{i-2} \right) }  \nonumber \\
&\times \frac{ e^{i \Omega \left( p^\prime, \vec{q}_{i,\perp} + ... + \vec{q}_{n,\perp} \right) \left( z^i - z^{i-1} \right) } - e^{i \Omega \left( p^\prime + k_\gamma, \vec{q}_{i,\perp} + ... + \vec{q}_{n,\perp} \right) \left(z^i - z^{i-1} \right) } }{ \Omega \left( p^\prime, \vec{q}_{i,\perp} + ... + \vec{q}_{n,\perp} \right) -  \Omega \left( p^\prime + k_\gamma, \vec{q}_{i,\perp} + ... + \vec{q}_{n,\perp} \right)} \nonumber \\
&\times e^{i \Omega \left( p^\prime, \vec{q}_{i+1,\perp} + ... + \vec{q}_{n,\perp} \right) \left(z^{i+1} - z^i \right) }  ... e^{i \Omega \left( p^\prime, \vec{q}_{n,\perp} \right) \left(z^n - z^{n-1} \right) }. \label{eq:amplitude_soft_n_order}
\end{align}
In the soft-photon factors of Eqs.~(\ref{eq:amplitude_soft_first_order})-(\ref{eq:amplitude_soft_n_order}), we neglected the change of momenta from the exchange of Glauber photons. We verified both analytically and numerically that the effects associated with this approximation are power suppressed.

We describe the evaluation of cross-section corrections in Appendix~\ref{app:derivation} and provide our results in the main text below. The resulting radiative cross-section corrections at first, second, and $n$th order of the opacity expansion $\delta \mathrm{d} \sigma_{1 \gamma}^{(1)}$, $\delta \mathrm{d} \sigma_{1 \gamma}^{(2)}$, and $\delta \mathrm{d} \sigma_{1 \gamma}^{(n)}$, respectively, are given by
\begin{align}
    \delta \mathrm{d} \sigma_{1 \gamma}^{(1)} &= e^2 \Bigg | \frac{ p^\prime \cdot \varepsilon^\star}{p^\prime \cdot k_\gamma} - \frac{ v \cdot \varepsilon^\star}{v \cdot k_\gamma} \Bigg |^2 \frac{\mathrm{d}^3 \vec{k}_\gamma}{\left(2 \pi \right)^3 2 E_\gamma} \sum \limits_{z^1 > 0} \frac{N_{z^1}}{S^{z^1}_\perp}  \int \frac{\mathrm{d}^2 \vec{q}_\perp}{\left( 2 \pi \right)^2}  {v} \left( {q}_{\perp} \right)^2  \nonumber \\
    & \times \left[  \mathrm{d} \sigma^\mathrm{LO} \left( \vec{p}^\prime + \vec{k}_\gamma - \vec{q}_\perp \right)  | \Gamma^{(1)} \left( \vec{q}_{\perp} , z^1 \right)  |^2   - \mathrm{d} \sigma^\mathrm{LO} \left( \vec{p}^\prime + \vec{k}_\gamma \right) \right], \label{eq:opacity_first_order} \\
    \delta \mathrm{d} \sigma_{1 \gamma}^{(2)} &= e^2 \Bigg | \frac{ p^\prime \cdot \varepsilon^\star}{p^\prime \cdot k_\gamma} - \frac{ v \cdot \varepsilon^\star}{v \cdot k_\gamma} \Bigg |^2 \frac{\mathrm{d}^3 \vec{k}_\gamma}{\left(2 \pi \right)^3 2 E_\gamma} \sum \limits_{z^2 > z^1 > 0} \frac{N_{z^1}}{S^{z^1}_\perp} \frac{N_{z^2}}{S^{z^2}_\perp}  \int \frac{\mathrm{d}^2 \vec{q}_{1,\perp}}{\left( 2 \pi \right)^2} \frac{\mathrm{d}^2 \vec{q}_{2,\perp}}{\left( 2 \pi \right)^2} {v} \left( q_{1,\perp} \right)^2 {v} \left( q_{2,\perp} \right)^2 \nonumber \\
    & \times \left[  \mathrm{d} \sigma^\mathrm{LO} \left( \vec{p}^\prime + \vec{k}_\gamma - \vec{q}_{1,\perp}- \vec{q}_{2,\perp} \right)  | \Gamma^{(2)} \left( \vec{q}_{1,\perp}, \vec{q}_{2,\perp}, z^1, z^2 \right)  |^2  + \mathrm{d} \sigma^\mathrm{LO} \left( \vec{p}^\prime + \vec{k}_\gamma \right) \right. \nonumber \\
    &\left. - \mathrm{d} \sigma^\mathrm{LO} \left( \vec{p}^\prime + \vec{k}_\gamma - \vec{q}_{1,\perp} \right)  | \Gamma^{(1)} \left( \vec{q}_{1,\perp}, z^1 \right)  |^2 - \mathrm{d} \sigma^\mathrm{LO} \left( \vec{p}^\prime + \vec{k}_\gamma - \vec{q}_{2,\perp} \right)  | \Gamma^{(1)} \left( \vec{q}_{2,\perp}, z^2 \right)  |^2   \right], \label{eq:opacity_second_order} \\
    \delta \mathrm{d} \sigma_{1 \gamma}^{(n)} &= e^2  \Bigg | \frac{ p^\prime \cdot \varepsilon^\star}{p^\prime \cdot k_\gamma} - \frac{ v \cdot \varepsilon^\star}{v \cdot k_\gamma} \Bigg |^2 \frac{\mathrm{d}^3 \vec{k}_\gamma}{\left(2 \pi \right)^3 2 E_\gamma} \sum \limits_{z^n > ... > z^1 > 0} \frac{N_{z^1}}{S^{z^1}_\perp} ... \frac{N_{z^n}}{S^{z^n}_\perp} \int \frac{\mathrm{d}^2 \vec{q}_{1,\perp}}{\left( 2 \pi \right)^2} ... \frac{\mathrm{d}^2 \vec{q}_{n,\perp}}{\left( 2 \pi \right)^2} {v} \left( q_{1,\perp} \right)^2 ... {v} \left( q_{n,\perp} \right)^2 \nonumber \\
    & \times \left[  \mathrm{d} \sigma^\mathrm{LO} \left( \vec{p}^\prime + \vec{k}_\gamma - \vec{q}_{\perp} \right)  | \Gamma^{(n)} \left( \vec{q}_{1,\perp}, ..., \vec{q}_{n,\perp}, z^1, ..., z^n \right)  |^2 \right. \nonumber \\
    &\left.  - \sum \limits_{i=1}^{n} \mathrm{d} \sigma^\mathrm{LO} \left( \vec{p}^\prime + \vec{k}_\gamma - \vec{q}_{\perp}  + \vec{q}_{i,\perp}  \right)  | \Gamma^{(n-1)} \left( \vec{q}_{1,\perp}, ..., \vec{q}_{i-1,\perp}, \vec{q}_{i+1,\perp}, ..., \vec{q}_{n,\perp}, z^1, ..., z^{i-1}, z^{i+1}, ..., z^n \right)  |^2  \right. \nonumber \\
    &\left.  + \sum \limits_{i=2}^{n}  \sum \limits_{j=1}^{i-1} \mathrm{d} \sigma^\mathrm{LO} \left( \vec{p}^\prime + \vec{k}_\gamma - \vec{q}_{\perp}  + \vec{q}_{i,\perp} + \vec{q}_{j,\perp}  \right)  \right. \nonumber \\
    &\left. \times | \Gamma^{(n-2)} \left( \vec{q}_{1,\perp}, ..., \vec{q}_{j-1,\perp}, \vec{q}_{j+1,\perp}, ..., \vec{q}_{i-1,\perp}, \vec{q}_{i+1,\perp}, ..., \vec{q}_{n,\perp}, z^1, ..., z^{j-1}, z^{j+1}, ..., z^{i-1}, z^{i+1}, ..., z^n \right)  |^2   \right. \nonumber \\
    &\left. + ... + \left( - 1 \right)^{n-1} \sum \limits_{i=1}^{n}  \mathrm{d} \sigma^\mathrm{LO} \left( \vec{p}^\prime + \vec{k}_\gamma - \vec{q}_{i,\perp}   \right)  | \Gamma^{(1)} \left( \vec{q}_{i,\perp}, z^i\right)  |^2 + \left( - 1 \right)^n  \mathrm{d} \sigma^\mathrm{LO} \left( \vec{p}^\prime + \vec{k}_\gamma \right) \right], \label{eq:opacity_n_order}
\end{align}
with $\vec{q}_{\perp} = \sum \limits_{i=1}^n \vec{q}_{i,\perp}$, $N_{z^i}$ scattering centers that are distributed in the cross-sectional area $S^{z^i}_\perp$ around the interaction point $z^i$ along the charged lepton trajectory, and the third component of the hard interaction space-time point $z^0 = 0$. The relative effect from each order of the opacity expansion without modification of phases by the radiation, i.e., $\Gamma^{(n)} = 1$, is equivalent to the corrections to radiation-free cross sections from the exchange of Glauber photons, cf. Ref.~\cite{Tomalak:2022kjd}. Results of Ref.~\cite{Tomalak:2022kjd} serve as a reference point for the first order of the opacity expansion. This relative cross-section correction does not depend on the details of photon kinematics.

\subsection{Soft-photon radiation to all orders in $\alpha$ at leading $\mathrm{SCET}_\mathrm{G}$ power} \label{sec:soft_all_orders}

At the leading power of $\mathrm{SCET}_\mathrm{G}$, all additional photon-induced factors are the same,\footnote{Contrary to the estimates in Ref.~\cite{Vitev:2008vk} that are based on the exchange of Glauber gluons with QCD medium, soft photons contribute beyond the first two orders in the opacity expansion of Glauber-photon interactions with QED medium, both at leading and subleading powers.}
\begin{equation}
    |\Gamma^{(1)}\left( \vec{q}_{\perp}, z^1 \right)|^2 = |\Gamma^{(2)} \left( \vec{q}_{1,\perp}, \vec{q}_{2,\perp}, z^1, z^2 \right)|^2 = ... = |\Gamma^{(n)} \left( \vec{q}_{1,\perp} ... \vec{q}_{n,\perp}, z^1 ... z^n \right)|^2 = 1.
\end{equation}
It allows us to resum all orders of the opacity expansion. At each order, we obtain the multiplication of the radiation-free expression with the eikonal factor and photon phase space. Integrating over the photon phase space, accounting for the soft-photon region of virtual diagrams, and including one soft photon of energy $E_\gamma$ below $\Delta E$, we describe the soft-photon bremsstrahlung with the soft function $S_{1\gamma} \left( \Delta E \right)$, cf. Refs.~\cite{Hill:2016gdf,Tomalak:2021hec,Tomalak:2022xup} for explicit expressions in electron- and (anti)neutrino-nucleon scattering. From general properties of radiation in QED in the infrared ~\cite{Yennie:1961ad,Hill:2016gdf,Tomalak:2021hec,Tomalak:2022xup}, double logarithms in the soft function can be resummed to account for the radiation of arbitrary number of photons with a total energy below $\Delta E$ as
\begin{align} \label{eq:soft_function}
  S_{1\gamma} \left(\Delta E \right) \to S \left(  \Delta E \right) = e^{S_{1\gamma} \left(\Delta E \right)-1}.
\end{align}
In the elastic scattering of the charged lepton off the nucleus, we combine a set of radiative corrections into the renormalization-scheme-independent one-photon soft function by adding virtual vertex correction, vacuum polarization, and soft-photon radiation from the initial and final charged lepton lines, which is given by the substitution of $v^\mu$ with the incoming lepton momentum $p^\mu$ in Eq.~(\ref{eq:soft_radiation_momenta_GI}). In the limit of large squared momentum transfers $Q^2 = - \left( p - p^\prime \right)^2  \gg m^2_\ell$~\cite{Maximon:2000hm,Vanderhaeghen:2000ws}, we obtain the gauge-invariant and scheme-independent soft function $S^\ell_{1\gamma}$
\begin{align}
  S^\ell_{1\gamma} = 1 + \frac{\alpha}{ \pi} \left( \ln \frac{\left(\Delta E\right)^2}{E_\ell E^\prime_\ell} \left[ \ln \frac{Q^2}{m^2_\ell} - 1 \right] - \frac{1}{2} \ln^2 \frac{E_\ell}{E^\prime_\ell}+ \frac{13}{6} \ln \frac{Q^2}{m^2_\ell} + \mathrm{Li}_2 \left[ \cos^2 \frac{\theta_\ell}{2} \right] - \frac{\pi^2}{6} - \frac{28}{9} \right),\label{eq:soft_function_modification}
\end{align}
with the initial- and final-lepton energies $E_\ell$ and $E_\ell^\prime$, respectively. In Eq.~(\ref{eq:soft_function_modification}), we consider only the electron vacuum polarization and do not account for other vertex structures that contribute the scheme- and gauge-independent constant terms, besides the Dirac one. To set the $\overline{\mathrm{MS}}$ renormalization scale for the soft function in (anti)neutrino-nucleus scattering, we notice that Eq.~(\ref{eq:soft_function_modification}) corresponds to the function from Ref.~\cite{Hill:2016gdf} at the $\overline{\mathrm{MS}}$ renormalization scale $\mu \sim \Delta E$.

\begin{figure}[!t]
    \centering
    \includegraphics[height=0.49\textwidth]{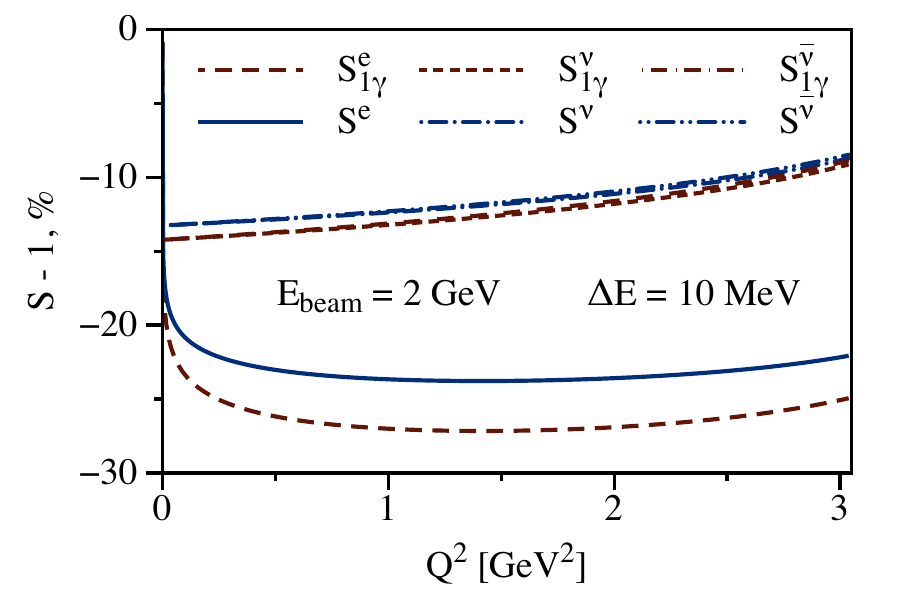}
    \caption{Soft-photon corrections in the elastic electron-, electron flavor neutrino-, and antineutrino-nucleus scattering are shown as a function of the squared momentum transfer for an incoming beam energy $E_\mathrm{beam} = 2~\mathrm{GeV}$ and the soft-photon energy cutoff $\Delta E = 10~\mathrm{MeV}$. Contributions from the one-photon soft functions in electron-, neutrino-, and antineutrino-nucleus scattering $S^e_{1\gamma},$ $S^\nu_{1\gamma},$ and $S^{\overline{\nu}}_{1\gamma}$, respectively, are compared to the exponentiated results $S^e,$ $S^\nu,$ and $S^{\overline{\nu}}$.} \label{fig:soft_suppression}
\end{figure}

In Fig.~\ref{fig:soft_suppression}, we illustrate the soft factors in electron-, electron flavor neutrino- and antineutrino-nucleus scattering for an incoming beam energy $E_\mathrm{beam} = 2~\mathrm{GeV}$ and a soft-photon energy cutoff $\Delta E = 10~\mathrm{MeV}$ as a function of the squared momentum transfer $Q^2$. The double-logarithmic contributions in elastic electron- and (anti)neutrino-nucleus scattering, which can be exponentiated as in Eq.~(\ref{eq:soft_function}), suppress cross sections at the $10\%$-$30\%$ level for electrons of GeV energies. Such contributions are important in precise measurements with electrons and neutrinos. The corrections from next-to-leading order in $\mathrm{SCET}_\mathrm{G}$ are smaller than the width of the lines in Fig.~\ref{fig:soft_suppression}.

Since the same soft factor arises at each order of the opacity expansion, it can be factored out and the Glauber-photon exchange series can be resummed for the lepton broadening distribution. The corresponding charged lepton $p^\prime_\perp$ distribution~\cite{Ovanesyan:2011xy} in the elastic (anti)neutrino-nucleus scattering is multiplied with the soft function $S$ as follows:
\begin{align}
\frac{\mathrm{d} N}{\mathrm{d} p^\prime_\perp} =  S \left( \Delta E  \right) \int \limits^{\infty}_{0} b p^\prime_\perp J_0 (0, b p^\prime_\perp) e^{- \chi \left[ 1 - \zeta b  K_1 \left( \zeta b \right) \right]} \mathrm{d} b, \label{eq:pT_distribution_soft}
\end{align}
with the modified Bessel function of the second kind $K_1$, the Bessel function of the first kind $J_0$, and an integral in the transverse coordinate space over the radial coordinate $b$. We observe the well-known double-logarithmic suppression of radiative corrections~\cite{Sudakov:1954sw} in both (anti)neutrino-nucleus and electron-nucleus scattering,  as it is illustrated in Fig.~\ref{fig:soft_suppression}. After averaging over the nuclear sites at each order of the opacity expansion, the multiple rescattering within the QED medium does not change the spectrum of radiated soft photons, but  rather slightly shifts the overall normalization, i.e., the cross section.

The soft-photon energy spectrum up to all orders of the opacity expansion at the leading power of $\mathrm{SCET_G}$ is easily obtained from the broadening distribution after integrating the derivative of Eq.~(\ref{eq:pT_distribution_soft}) with respect to the soft-photon energy cutoff $\Delta E$ over the $p_\perp^\prime$ spectrum. In the small lepton-mass limit, $m^2_\ell \ll Q^2, \tilde{E}^2_\mathrm{\ell}$, the bremsstrahlung spectrum can be expressed as\footnote{All the dependence on the longitudinal final-state lepton momentum cancels since the distribution is normalized to the tree-level cross section.}
\begin{align}
 \frac{\mathrm{d} N^{\gamma}_S}{\mathrm{d} x} =   \frac{\alpha}{\pi} \frac{S \left( E_\gamma \right)}{x}  \left(  \delta_\ell \ln \frac{Q^2}{m^2_\ell} + \frac{\delta_{\nu_\ell}}{2} \ln \frac{4 \tilde{E}_\ell^2}{m^2_\ell} - 1\right),
\end{align}
with the fraction of the photon energy in the sum of the charged lepton and photon energies $x = E_\gamma / \left( E_\ell^\prime + E_\gamma\right)$. Here, $\delta_\ell = 0,~\delta_{\nu_\ell} = 1$ for the (anti)neutrino scattering, and $\delta_\ell = 1, \delta_{\nu_\ell} = 0$ for the charged lepton scattering; $\tilde{E}_\ell$ denotes the charged lepton energy in the rest frame of the struck or final-state proton. This spectrum coincides with the soft-photon spectrum in vacuum. Consequently, the medium-induced radiative energy spectrum of soft photons on top of the vacuumlike one at leading $\mathrm{SCET}_\mathrm{G}$ power vanishes after the resummation of all orders of the opacity expansion.\footnote{The power-suppressed contributions from Ref.~\cite{Vitev:2008vk} are not necessary zero.}

\section{Radiation of collinear photons} \label{sec:collinear_section}

In this Section, we evaluate cross sections for the radiation of collinear photons. We derive analytic expressions for the first three orders in the opacity expansion and present the generalization to higher orders. We further study the soft-photon, hard-photon, and phase-independent limits, and discuss the generalization to higher orders and resummation of all orders. We also compare theoretical aspects of our calculation to the corresponding details of the derivation in QCD.

\subsection{Opacity expansion of the collinear radiation} \label{sec:collinear_opacity}

First, we consider the medium-induced radiation of one real collinear photon and evaluate cross-section corrections at each order of the opacity expansion.

The radiation of one collinear photon from the charged lepton multiplies the tree-level matrix element $\Bigg | M^\mathrm{LO}_{1 \gamma} \left( p^\prime + k_\gamma \right) \Bigg |^2$ as
\begin{align}
\mathrm{d} \sigma_{1 \gamma}^\mathrm{LO} = e^2 \frac{4}{{A}^2_{\perp,1}} \frac{1 +  \left( 1 - x \right)^2}{2} \left( 1 - x \right) \Bigg | M^\mathrm{LO}_{1 \gamma} \left( p^\prime + k_\gamma \right) \Bigg |^2 \frac{\mathrm{d}^3 \vec{p}^\prime}{ \left( 2 \pi \right)^3 2 E^\prime} \frac{\mathrm{d}^3 \vec{k}_\gamma}{ \left( 2 \pi \right)^3 2 E_\gamma}, \label{eq:collinear_radiation}
\end{align}
with the vector $\vec{A}_{\perp,1} = \vec{k}_{\gamma,\perp} (1-x) - \vec{p}^\prime_\perp x $, transverse to the direction of propagation, and fraction of the photon energy in the sum of lepton and photon energies $x = k_\gamma^+/ \left( p^\prime + k_\gamma \right)^+$. For the choice of the collinear direction along the sum of the photon and lepton momenta, i.e., $\vec{p}^\prime_\perp + \vec{k}_{\gamma,\perp} =0$ , we obtain the electron-photon splitting function $P_{e \to e} \left( x \right)$, analogous to the Altarelli-Parisi quark-gluon splitting~\cite{Gribov:1972ri,Gribov:1972rt,Lipatov:1974qm,Dokshitzer:1977sg,Altarelli:1977zs,Dokshitzer:1978hw,Peskin:1995ev}:
\begin{align}
\frac{\mathrm{d} \sigma_{1 \gamma}^\mathrm{LO}}{ \mathrm{d} \sigma^\mathrm{LO} \left( \vec{p}^\prime + \vec{k}_\gamma \right)} \frac{2 \pi \vec{k}^2_{\gamma,\perp}}{\mathrm{d} x \mathrm{d}^2 \vec{k}_{\gamma,\perp}}  = \frac{\alpha}{\pi} P_{e \to e} \left( x \right) = \frac{\alpha}{\pi} \frac{1 +  \left( 1 - x \right)^2}{x}, \label{eq:collinear_radiation_splitting}
\end{align}
where the radiation-free cross section for the tree level is expressed in terms of the electromagnetic instead of the electron kinematics. We have also verified the tree-level $\gamma \to e^+ e^-$ splitting function $P_{\gamma \to e} \left( x \right)$: $P_{\gamma \to e} \left( x \right) = x^2 + \left( 1 - x \right)^2$, where the electron carries a fraction of the photon energy $x$. In this work, we regularize the collinear divergence by introducing the nonzero charged lepton mass through the velocity parameter $\beta_\ell$ and obtain the energy spectrum of collinear photons,
\begin{align}
\frac{\mathrm{d} N^\gamma_c}{\mathrm{d} x} &= \frac{\mathrm{d} \sigma_{1 \gamma}^\mathrm{LO}}{ \mathrm{d} \sigma^\mathrm{LO} \left( \vec{p}^\prime + \vec{k}_\gamma \right) \mathrm{d} x } \equiv \frac{\alpha}{2 \pi} \frac{1 +  \left( 1 - x \right)^2}{x} \frac{\beta_\ell \sin \theta_\gamma }{1-\beta_\ell \cos \theta_\gamma} \mathrm{d} \theta_\gamma, \label{eq:collinear_radiation_regularization} \\
\frac{\mathrm{d} N^\gamma_c}{\mathrm{d} x} &\approx \frac{\alpha}{\pi} \frac{1 +  \left( 1 - x \right)^2}{x} \ln \frac{\left( E^\prime_\ell + E_\gamma \right) \Delta \theta}{m_\ell}, \label{eq:collinear_radiation_regularization2}
\end{align}
where we consider the jet around the charged lepton direction with the cone size $\Delta \theta$, denote the angle between the electron and photon as $\theta_\gamma$, and keep only the leading power in $m_\ell$. The splitting-function approximation works only outside the dead cone when $ \left( E^\prime_\ell + E_\gamma \right) \theta_\gamma / m_\ell \gg 1$. Note that our radiative cross section is normalized to the leading-order cross section, which we evaluate with the electromagnetic $E_\ell^\prime + E_\gamma$ instead of the electron energy $E_\ell^\prime$.

The derivation of cross sections from matrix elements follows the same steps as in Sec.~\ref{sec:soft_opacity} for the soft-photon radiation from the charged lepton line and results in the same expressions as in Eqs.~(\ref{eq:opacity_first_order})-(\ref{eq:opacity_n_order}) with the only change of the soft-photon prefactor $e^2 \Big | \frac{ p^\prime \cdot \varepsilon^\star}{p^\prime \cdot k_\gamma} - \frac{ v \cdot \varepsilon^\star}{v \cdot k_\gamma} \Big |^2$ to the prefactors of the collinear radiation in vacuum, cf. Eqs.~(\ref{eq:collinear_radiation})-(\ref{eq:collinear_radiation_regularization}), and change in the photon-induced phase-dependent functions $\Gamma^{(n)}$ as
\begin{align} 
|\Gamma^{(1)} |^2  &= 1 + 2\frac{{A}^2_{\perp,1}}{A^2_{\perp,2}}  \left( 1 - \cos \frac{A^2_{\perp,2} \left( z^1 - z^0 \right)}{x \left(p^\prime \right)^+}   \right) - 2 \frac{\vec{A}_{\perp,1} \cdot \vec{A}_{\perp,2}}{A^2_{\perp,2}} \left( 1 - \cos \frac{A^2_{\perp,2} \left( z^1 - z^0 \right)}{x \left(p^\prime \right)^+} \right), \label{eq:opacity_first_order_collinear} \\
|\Gamma^{(2)} |^2  &= 1 + 2 \frac{{A}^2_{\perp,1}}{A^2_{\perp,3}}  \left( 1 - \cos \frac{A^2_{\perp,3} \left( z^1 - z^0 \right)}{x \left(p^\prime \right)^+}   \right) + 2\frac{{A}^2_{\perp,1}}{A^2_{\perp,2}} \left( 1 - \cos \frac{A^2_{\perp,2} \left( z^2 - z^1 \right)}{x \left(p^\prime \right)^+}   \right) \nonumber \\
&-  2   \frac{{A}^2_{\perp,1}}{A^2_{\perp,3}} \frac{\vec{A}_{\perp,2} \cdot \vec{A}_{\perp,3}}{A^2_{\perp,2}}  \left(  \cos \left[ \frac{A^2_{\perp,3} \left( z^1 - z^0 \right) + A^2_{\perp,2} \left( z^2 - z^1 \right)}{x \left(p^\prime \right)^+}  \right] - \cos \frac{A^2_{\perp,2} \left( z^2 - z^1 \right)}{x \left(p^\prime \right)^+}  \right)\nonumber \\
&-  2   \frac{{A}^2_{\perp,1}}{A^2_{\perp,3}} \frac{\vec{A}_{\perp,2} \cdot \vec{A}_{\perp,3}}{A^2_{\perp,2}}  \left( 1 - \cos \frac{A^2_{\perp,3} \left( z^1 - z^0 \right)}{x \left(p^\prime \right)^+}   \right)\nonumber \\
&-  2  \frac{\vec{A}_{\perp,1} \cdot \vec{A}_{\perp,3}}{A^2_{\perp,3}} \left( \cos \frac{A^2_{\perp,2} \left( z^2 - z^1 \right)}{x \left(p^\prime \right)^+}  - \cos \left[ \frac{A^2_{\perp,3}\left( z^1 - z^0 \right) + A^2_{\perp,2} \left( z^2 - z^1 \right)}{x \left(p^\prime \right)^+}  \right] \right)\nonumber \\
&- 2 \frac{\vec{A}_{\perp,1} \cdot \vec{A}_{\perp,2}}{A^2_{\perp,2}} \left(  1 - \cos \frac{A^2_{\perp,2} \left( z^2 - z^1 \right)}{x \left(p^\prime \right)^+} \right), \label{eq:opacity_second_order_collinear} \\
|\Gamma^{(n)} |^2 &= 1 + 2 {A}^2_{\perp,1} \sum \limits_{i=2}^{n+1} \sum \limits_{j=2}^{i} \frac{\vec{A}_{\perp, i} \cdot \vec{A}_{\perp, j}}{A^2_{\perp, i} A^2_{\perp, j}} \left(\cos \frac{\sum \limits_{k=j}^{i-1} {A}^2_{\perp, k} \left( z^{n + 2 - k} - z^{n + 1 - k} \right) }{x \left(p^\prime \right)^+}- \cos \frac{ \sum \limits_{k=j}^{i} {A}^2_{\perp, k} \left( z^{n + 2 - k} - z^{n + 1 - k} \right) }{x \left(p^\prime \right)^+}   \right) \nonumber \\
&+ 2 {A}^2_{\perp,1} \sum \limits_{i=2}^{n+1} \sum \limits_{j=1}^{i-1} \frac{\vec{A}_{\perp, i} \cdot \vec{A}_{\perp, j}}{A^2_{\perp, i} A^2_{\perp, j}} \left(\cos \frac{\sum \limits_{k=j+1}^{i} {A}^2_{\perp, k} \left( z^{n + 2 - k} - z^{n + 1 - k} \right) }{x \left(p^\prime \right)^+} - \cos \frac{ \sum \limits_{k=j+1}^{i-1} {A}^2_{\perp, k} \left( z^{n + 2 - k} - z^{n + 1 - k} \right) }{x \left(p^\prime \right)^+}    \right), \label{eq:opacity_n_order_collinear}
\end{align}
with the transverse vectors
\begin{align}
\vec{A}_{\perp,i} &= \vec{k}_{\gamma,\perp} \left(1-x\right) - \left( \vec{p}^\prime_\perp - \vec{q}_{1,\perp} - ... - \vec{q}_{i-1,\perp} \right) x, \qquad i = 1, ..., n.
\end{align}
We replace the sum with zero when the lowest limit in the sum exceeds the upper limit. Contrary to the soft radiation, the collinear radiation from the incoming charged lepton does not interfere with the bremsstrahlung from the outgoing lepton for sufficiently large scattering angles. In this work, we do not consider the collinear radiation for the forward kinematics when the interference should be taken into account.

For completeness, we also provide the function $|\Gamma^{(1)} |^2$ for the leading in the opacity QED nuclear medium-induced contribution to $\gamma \to e^+ e^-$ splitting function,
\begin{align}
     |\Gamma^{(1)} |^2 &= 2 \cos \left[ \left( - \frac{q^2_\perp}{\left( k^+_\gamma \right)^2} + 2 \left( 1 - 2 x \right)    \frac{\vec{p}^e_\perp \cdot \vec{q}_\perp}{\left( k^+_\gamma \right)^2}\right) \frac{z^1 - z^0}{x \left( 1 - x \right) } \right]- 1,
\end{align}
with the same form of the cross-section correction as for the soft and collinear radiation.

\subsection{Collinear radiation to all orders in $\alpha$ at leading $\mathrm{SCET}_\mathrm{G}$ power} \label{sec:collinear_all_orders}

Setting first $\vec{A}_{\perp,1} = \vec{A}_{\perp,2} = ... = \vec{A}_{\perp,n}$ only in the numerators, but not in the denominators and not under the arguments of the trigonometric functions, and, subsequently, $x=0$, we reproduce the result for the medium-induced bremsstrahlung correction of soft photons from Sections~\ref{sec:soft_opacity} and~\ref{sec:soft_all_orders} that multiplies the collinear enhancement in vacuum.

It is instructive to consider two more limits: 1) the limit $x \to 1$ with rapidly oscillating arguments of trigonometric functions, and 2) the limit of small trigonometric arguments that corresponds to the typical experimental situation and phase-independent functions $\Gamma^{(n)}$. In both limits, the first term, $1$, from the $\Gamma^{(n)} = 1 + ...$ functions results in the same relative nuclear medium correction that multiplies the collinear enhancement and the splitting function in vacuum in the same way as for the soft radiation and as for the radiation-free process. We give the analytical derivation of these limits in Appendix~\ref{app:limit_large_x} for limit 1 and in Appendix~\ref{app:limit_intermediate_x} for the limit 2, respectively. We find that the leading nuclear medium-induced correction to the bremsstrahlung is determined by kinematic changes of cross sections in Eqs.~(\ref{eq:opacity_first_order})-(\ref{eq:opacity_n_order}) and generalize this result to all orders, i.e., $\Gamma^{(n)} = 1$ determines not only soft and radiation-free QED nuclear modifications but also nuclear medium-induced contributions to the collinear radiation. By this, we find that the medium-induced radiative energy spectrum of collinear photons is vacuumlike and any corrections vanish after the resummation of all orders of the opacity expansion, while the associated charged lepton broadening distribution is multiplied by the corresponding jet function in vacuum~\cite{Tomalak:2021hec,Tomalak:2022xup}. The perturbative resummation of double-logarithmic corrections can be performed by the exponentiation of the leading-order double logarithms.

\subsection{Collinear radiation at the first order of the opacity expansion} \label{sec:collinear_first_opacity_order_numerics}

In our estimates below, we verify the conjecture $\Gamma^{(1)} \approx 1$ numerically for the first order of the opacity expansion and reproduce the QED nuclear medium correction of Ref.~\cite{Tomalak:2022kjd}.

In addition to the kinematics of the process, the nuclear medium-induced spectral distortions depend on two dimensionless variables $\frac{p^\prime_\perp}{\zeta}$ and $\frac{p^\prime_\perp}{E^\prime_\ell + E_\gamma} \times p^\prime_\perp R_\mathrm{rms}$, with the nuclear size $R_\mathrm{rms}$ and overall prefactor $1/\left( \zeta R_\mathrm{rms} \right)^2$. Regulating the collinear divergence with the finite lepton mass and exploiting the kinematic relation at leading power $p^\prime_\perp = \frac{ E^\prime_\ell E_\gamma}{E^\prime_\ell +E_\gamma } \theta_\ell$, we integrate over the photon angles and obtain the spectrum of collinear radiation within a cone of some size $\Delta \theta$. In the following Figs.~\ref{fig:radiation_collinear_neutrino_005}-\ref{fig:radiation_collinear_electron_03}, we study the spectral distortions of radiated collinear photons in elastic neutrino-neutron, antineutrino-proton, and electron-nucleon scattering inside the nucleus, including averaging over the nuclear volume. We present the nuclear medium-induced contribution at the first order of the opacity expansion for the incoming beam energy $2~\mathrm{GeV}$, the cone size $\Delta \theta = 10^\mathrm{o}$, and the squared momentum transfers $0.05~\mathrm{GeV}^2$ and $0.3~\mathrm{GeV}^2$. Our numerical results reproduce the soft-photon limit $x \to 0$ of the $x$-independent ratio of the nuclear medium-induced radiation to the bremsstrahlung in vacuum. The resulting ratio is the same as the relative QED nuclear medium-induced corrections to the radiation-free process~\cite{Tomalak:2022kjd}. This ratio is positive for scattering of antineutrinos at $Q^2 = 0.3~\mathrm{GeV}^2$ and electrons at any squared momentum transfer and negative for scattering of neutrinos and antineutrinos at $Q^2 = 0.05~\mathrm{GeV}^2$ and neutrinos at both $Q^2 = 0.05~\mathrm{GeV}^2$ and $Q^2 = 0.3~\mathrm{GeV}^2$ squared momentum transfers. Compared to the soft radiation in the vacuum, cf. Fig.~\ref{fig:soft_suppression}, the collinear nuclear medium-induced radiation at the first order of the opacity expansion is suppressed by orders of magnitude. As it is expected and represented in Figs.~\ref{fig:radiation_collinear_neutrino_005}-\ref{fig:radiation_collinear_electron_03}, the nuclear medium-induced radiation increases with the size of the nucleus.
\begin{figure}[t]
    \centering
    \includegraphics[height=0.33\textwidth]{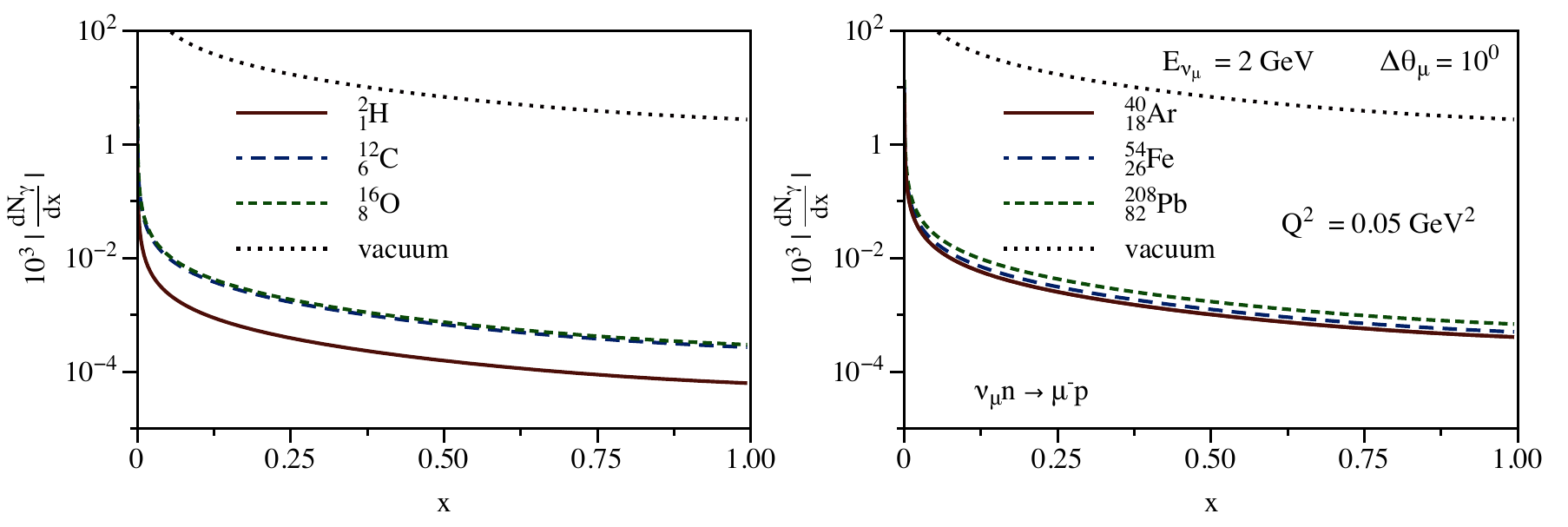}
    \caption{The energy spectrum of QED nuclear medium-induced collinear photons is compared to the collinear radiation in vacuum for the charged-current elastic neutrino scattering process. The absolute value for the distribution corresponds to the negative contribution of the QED nuclear medium-induced bremsstrahlung to the total photon energy spectrum. The incoming muon neutrino energy is $E_{\nu_\mu} = 2~\mathrm{GeV}$ and the squared momentum transfer is $Q^2 = 0.05~\mathrm{GeV}^2$. The relative QED nuclear medium-induced correction does not depend on $x$.} \label{fig:radiation_collinear_neutrino_005}
\end{figure}
\begin{figure}[t]
    \centering
    \includegraphics[height=0.33\textwidth]{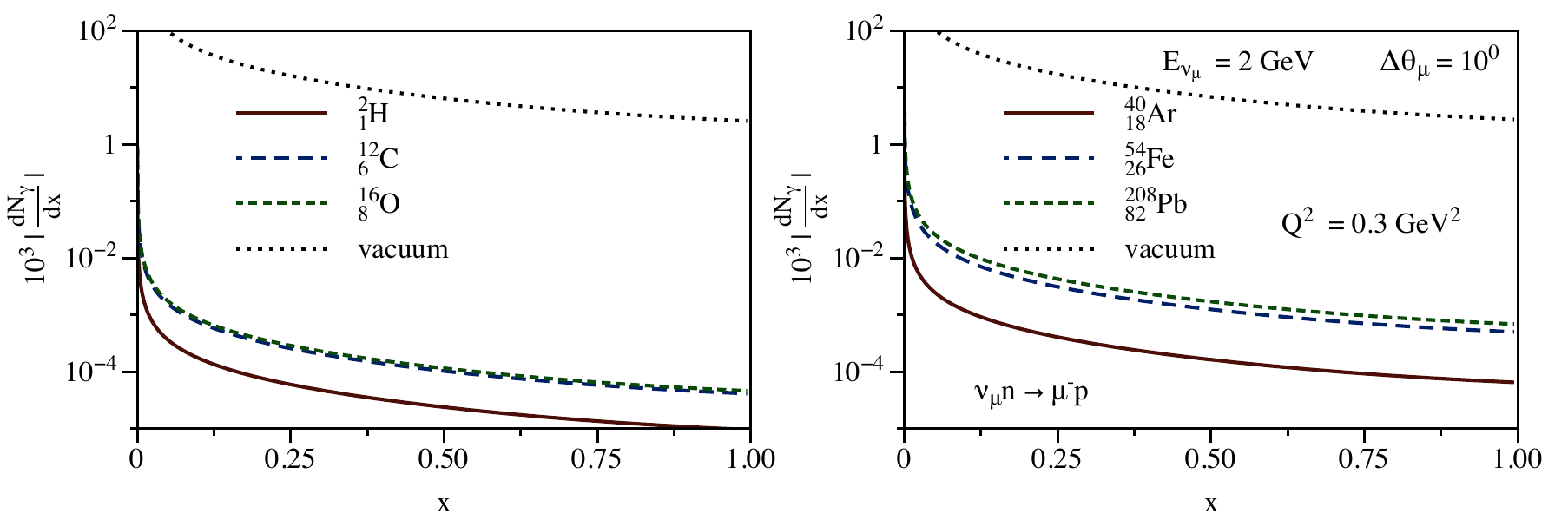}
    \caption{Same as Fig.~\ref{fig:radiation_collinear_neutrino_005}, but for the squared momentum transfer $Q^2 = 0.3~\mathrm{GeV}^2$.} \label{fig:radiation_collinear_neutrino_03}
\end{figure}
\begin{figure}[t]
    \centering
    \includegraphics[height=0.33\textwidth]{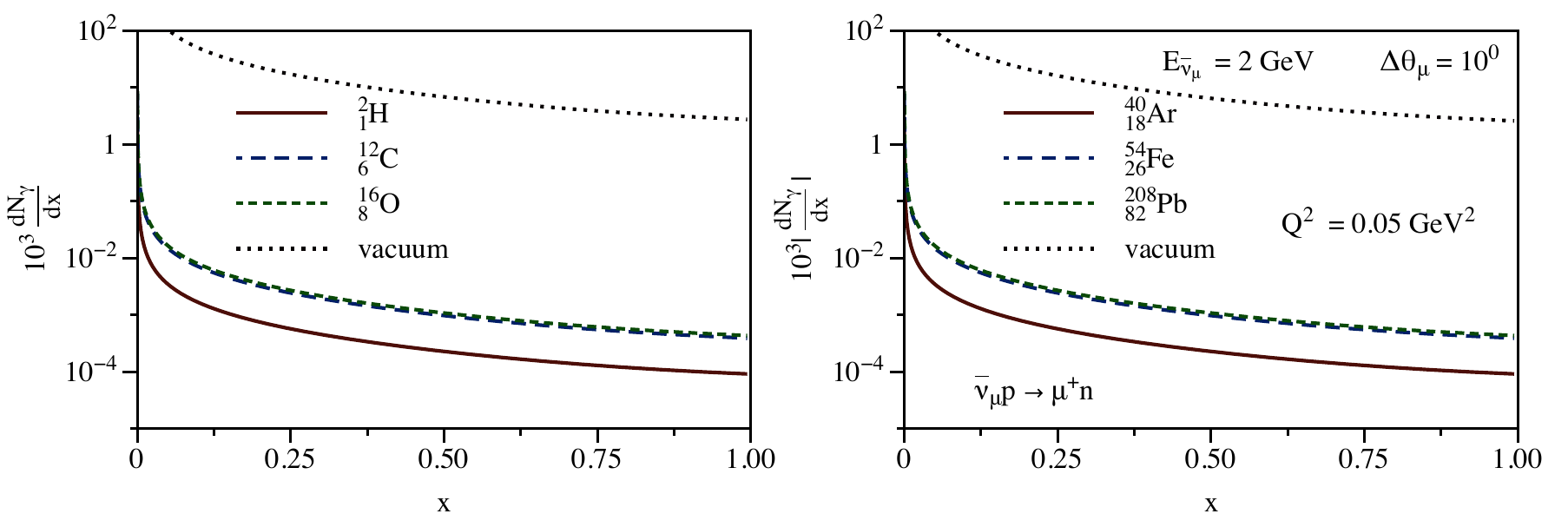}
    \caption{Same as Fig.~\ref{fig:radiation_collinear_neutrino_005}, but for the antineutrino scattering. We do not indicate the absolute value for positive contributions to the photon energy spectrum.} \label{fig:radiation_collinear_antineutrino_005}
\end{figure}
\begin{figure}[t]
    \centering
    \includegraphics[height=0.33\textwidth]{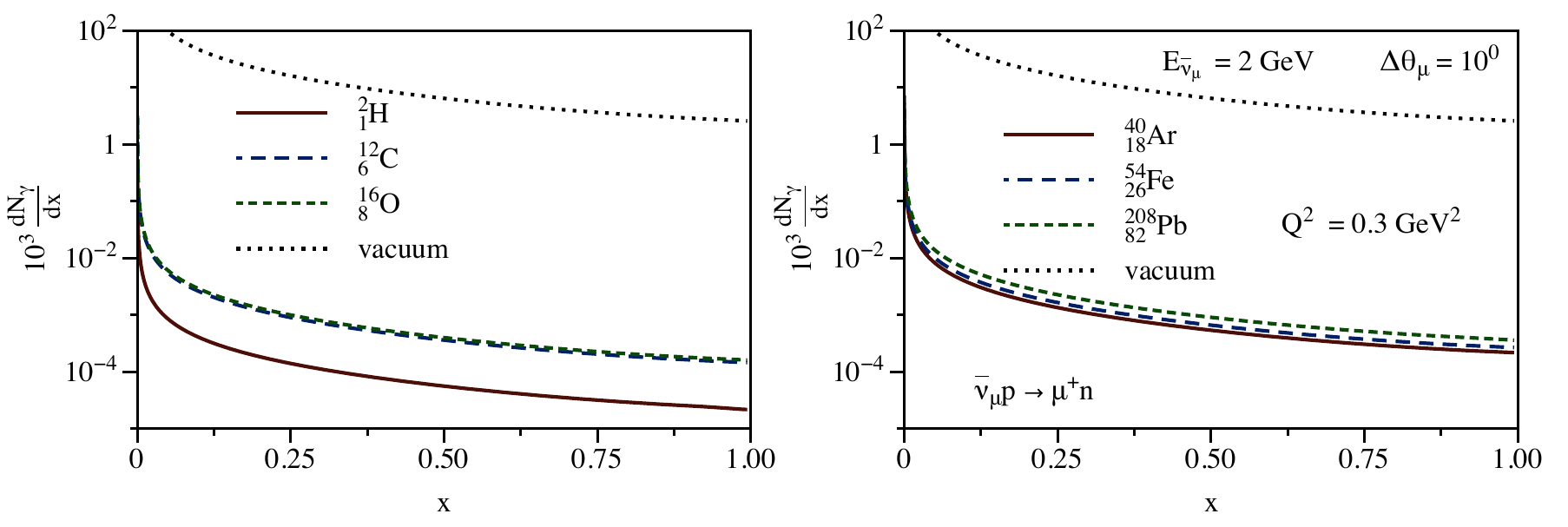}
    \caption{Same as Fig.~\ref{fig:radiation_collinear_neutrino_03}, but for the antineutrino scattering.} \label{fig:radiation_collinear_antineutrino_03}
\end{figure}
\begin{figure}[t]
    \centering
    \includegraphics[height=0.33\textwidth]{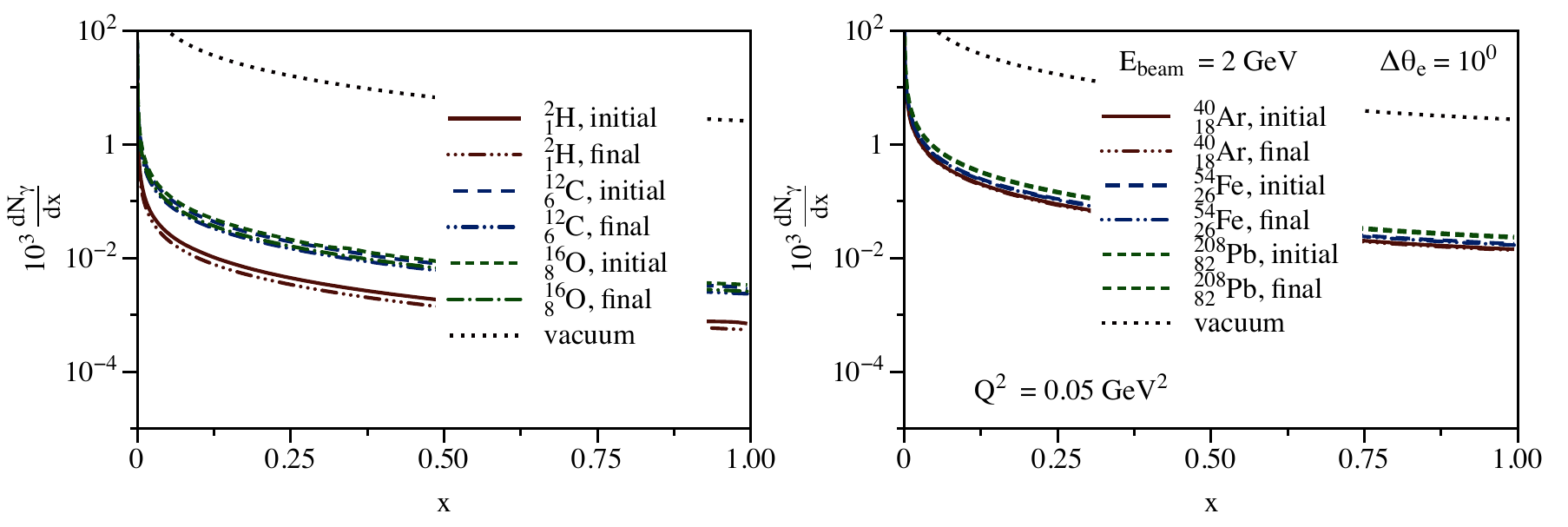}
    \caption{Same as Fig.~\ref{fig:radiation_collinear_neutrino_005}, but for the electron scattering. Initial- and final-state photon energy spectra are shown.} \label{fig:radiation_collinear_electron_005}
\end{figure}
\begin{figure}[t]
    \centering
    \includegraphics[height=0.33\textwidth]{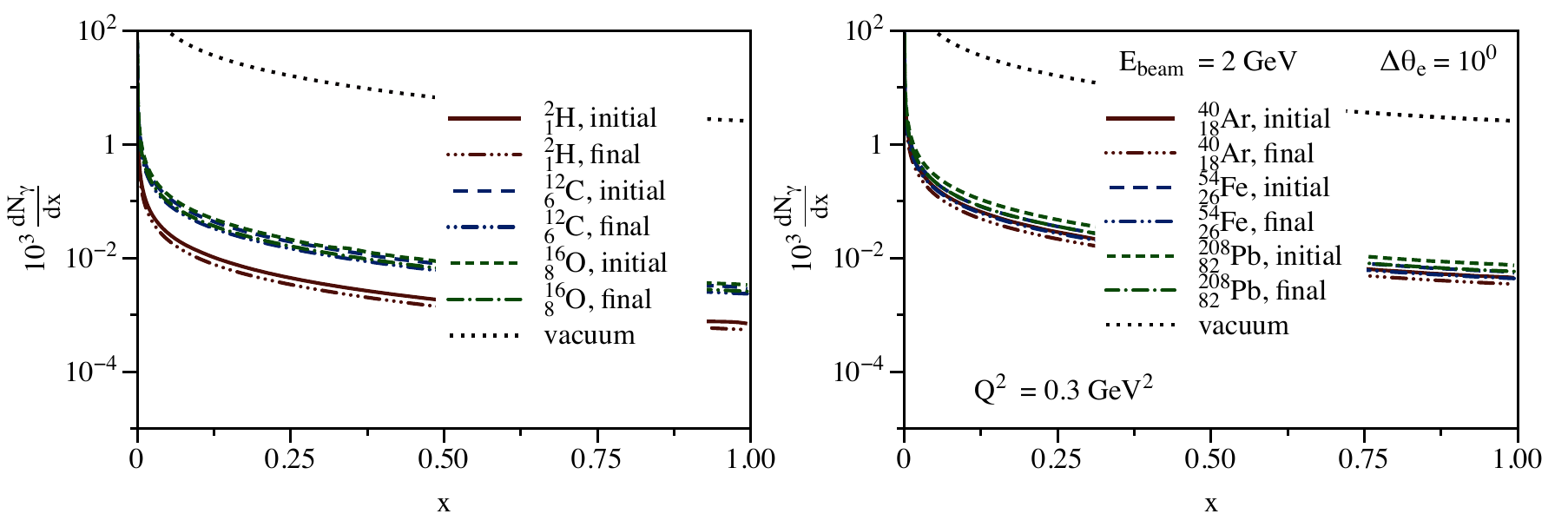}
    \caption{Same as Fig.~\ref{fig:radiation_collinear_neutrino_03}, but for the electron scattering. Initial- and final-state photon energy spectra are shown.} \label{fig:radiation_collinear_electron_03}
\end{figure}

\subsection{Comparison to radiation in QCD} \label{sec:collinear_QCD}

The tree-level radiation of soft and collinear gluons in QCD~\cite{Ovanesyan:2011xy} is obtained by multiplying the QED result with the one-loop color factor $C_F$. Already at the first order of the opacity expansion, there is no direct relation between QCD and QED due to the presence of gluon self-interaction in QCD and vanishing of amplitudes with the one color factor in quark-gluon vertices~\cite{Tomalak:2023kwl}.

\newpage
\section{Radiative energy loss in vacuum and inside the nucleus} \label{sec:energy_loss}

In this Section, we apply the radiative energy-loss formalism of Refs.~\cite{Gyulassy:2000er,Gyulassy:2001nm} and the bremsstrahlung spectra from Secs.~\ref{sec:soft_section} and~\ref{sec:collinear_section} for the evaluation of QED nuclear medium effects on the kinematics of the outgoing charged lepton in the (anti)neutrino-nucleus scattering and both incoming and outgoing electrons in the elastic electron-nucleus scattering.

We consider four types of the bremsstrahlung: the nuclear medium-induced radiation of one soft ($N^1_S$) or collinear ($N^1_c$) photon at the first order of the opacity expansion, the radiation of one soft photon at all orders in the opacity expansion at leading order in $\alpha$ ($N^{1\gamma}_S$), and the radiation of many soft photons after the resummation of leading logarithms ($N^\gamma_S$). The corresponding photon energy spectra from Secs.~\ref{sec:soft_section} and~\ref{sec:collinear_section} can be expressed as
\begin{align}
    \frac{\mathrm{d} N^1_S}{\mathrm{d} x} &= \frac{\alpha}{\pi}  \frac{1}{x}  \left( \delta_\ell \ln \frac{Q^2}{m^2_\ell} + \frac{\delta_{\nu_\ell}}{2} \ln \frac{4 \tilde{E}_\ell^2}{m^2_\ell} - 1 \right) \sum \limits_{z^1 > 0} \frac{N_{z^1}}{S^{z^1}_\perp}  \int \frac{\mathrm{d}^2 \vec{q}_\perp}{\left( 2 \pi \right)^2}  {v} \left( {q}_{\perp} \right)^2  \left[  \frac{\mathrm{d} \sigma_{1 \gamma}^\mathrm{LO} \left( \vec{p}^\prime + \vec{k}_\gamma - \vec{q}_\perp \right)}{\mathrm{d} \sigma_{1 \gamma}^\mathrm{LO} \left( \vec{p}^\prime + \vec{k}_\gamma \right)}  - 1 \right], \\
   \frac{\mathrm{d} N^1_c}{\mathrm{d} x} &=  \frac{\alpha}{2 \pi} \frac{1 +  \left( 1 - x \right)^2}{x} \int \limits_{0}^{\Delta \theta} \frac{\beta_\ell \sin \theta_\gamma }{1-\beta_\ell \cos \theta_\gamma} \mathrm{d} \theta_\gamma \sum \limits_{z^1 > 0} \frac{N_{z^1}}{S^{z^1}_\perp } \int \frac{\mathrm{d}^2 \vec{q}_\perp}{\left( 2 \pi \right)^2}  {v} \left( {q}_{\perp} \right)^2 \nonumber \\
   &  \left[  \frac{\mathrm{d} \sigma_{1 \gamma}^\mathrm{LO} \left( \vec{p}^\prime + \vec{k}_\gamma - \vec{q}_\perp \right)}{\mathrm{d} \sigma_{1 \gamma}^\mathrm{LO} \left( \vec{p}^\prime + \vec{k}_\gamma \right)} | \Gamma^{(1)} \left( \vec{q}_\perp , z^1 \right)  |^2  - 1 \right], \\
   \frac{\mathrm{d} N^1_c}{\mathrm{d} x} &\approx \frac{\mathrm{d} N^1_S}{\mathrm{d} x} \left( 1 +  \left( 1 - x \right)^2 \right) \ln \frac{\left(E_\ell + E_\gamma \right) \Delta \theta}{m_\ell}  \left(  \delta_\ell \ln \frac{Q^2}{m^2_\ell} + \frac{\delta_{\nu_\ell}}{2} \ln \frac{4 \tilde{E}_\ell^2}{m^2_\ell} - 1\right)^{-1} , \\
   \frac{\mathrm{d} N^{1\gamma}_S}{\mathrm{d} x} &= \frac{\alpha}{\pi}  \frac{1}{x} \left( \delta_\ell \ln \frac{Q^2}{m^2_\ell} + \frac{\delta_{\nu_\ell}}{2} \ln \frac{4 \tilde{E}_\ell^2}{m^2_\ell} - 1 \right), \\
   \frac{\mathrm{d} N^\gamma_S}{\mathrm{d} x} &= \frac{\alpha}{\pi}  \frac{S \left(E_\gamma \right)}{x}  \left(  \delta_\ell \ln \frac{Q^2}{m^2_\ell} + \frac{\delta_{\nu_\ell}}{2} \ln \frac{4 \tilde{E}_\ell^2}{m^2_\ell} - 1\right).
\end{align}

From these distributions, we can evaluate the mean number of the radiated photons $<N^\gamma>$ and the energy in the photon component $\Delta E_\gamma$ as
\begin{align}
    <N^\gamma> &= \int \mathrm{d} x \frac{\mathrm{d} N^\gamma}{\mathrm{d} x}, \\
    \Delta E_\gamma &=  \int \mathrm{d} x x \left( E_\ell + E_\gamma \right)  \frac{\mathrm{d} N^\gamma}{\mathrm{d} x}.
\end{align}
We extrapolate the soft-photon energy spectra outside the region of validity of the soft-photon calculation and obtain by this only an approximate estimate. To avoid ``infrared catastrophe", it makes sense to consider the mean number of radiated photons only for the photon energy above some cutoff. Considering an incoming lepton of energy $2~\mathrm{GeV}$ and squared momentum transfers $Q^2=0.05,~0.3,$ and~$1.2~\mathrm{GeV}^2$, we present the results for the lepton energy loss in Table~\ref{tab:energy_losses}. The radiative energy loss in vacuum is large, but it is included in the definition of kinematics in the evaluation of radiative corrections~\cite{Tomalak:2021hec,Tomalak:2022xup}. The fixed-order one-photon and resummed results for the radiation of soft photons are the same within a few significant digits. The addition to the vacuum QED nuclear medium-induced energy loss increases with the size of the nucleus and does not exceed $70~\mathrm{keV}$ for $^2_1\mathrm{H}$ and $600~\mathrm{keV}$ for $^{208}_{82}\mathrm{Pb}$. Having the energy loss below $1~\mathrm{MeV}$ for typical nuclei in neutrino- and electron-nucleus scattering, which is below the experimental resolution, we can safely neglect such an effect in the analysis of the experimental data. At the first order of the opacity expansion, we find similar energy loss that is caused by the radiation of soft and collinear photons within a $\Delta \theta = 10^\circ$ cone in (anti)neutrino-nucleus scattering, while the collinear energy loss in electron-nucleus scattering is an order of magnitude below the energy loss from soft photons. The bremsstrahlung contributions and energy losses from initial- and final-state electrons have similar size.
\begin{center}
\begin{table}[htb]
 \centering 
 \begin{tabular}{|c|c|c|c|c|c|}
  \hline
  & $\Delta E^{1\gamma}_S(\mathrm{MeV})$ & $\Delta E^{\gamma}_S(\mathrm{MeV})$ & $\Delta E^{1 \gamma}_c(\mathrm{MeV})$ & $\Delta E^{1}_S(\mathrm{MeV})$ & $\Delta E^{1}_c(\mathrm{MeV})$  \\ \hline
 $Q^2 = 0.05~\mathrm{GeV}^2$ &  &  &  &  &  \\ \hline
 $\nu_\mu$ & $12.2$ & $12.2$ & $7.3$ & $\left(0.2-2\right) \times 10^{-3}$ & $\left(0.2-2\right) \times 10^{-3}$ \\
 $\overline{\nu}_\mu$ &  $12.2$  & $12.2$ & $7.3$ & $ \left(0.3-2\right) \times 10^{-3}$ & $\left(0.2-3\right) \times 10^{-3}$ \\
 $\mathrm{initial}~{e}$ & $51.9$  & $51.9$ & $40.4$ & $0.07-0.6$ & $0.006-0.06$ \\
 $\mathrm{final}~{e}$ & $51.9$  & $51.9$ & $40.3$ & $0.07-0.6$ & $0.006-0.06$ \\ \hline
 $Q^2 = 0.3~\mathrm{GeV}^2$ &  &  &  &  &  \\ \hline
 $\nu_\mu$ & $11.8$  & $11.9$ & $6.9$ & $\left(0.3-3\right)\times10^{-4}$ & $\left(0.3-3\right)\times10^{-4}$  \\
 $\overline{\nu}_\mu$ &  $11.8$  & $11.9$ & $6.9$ & $\left(0.9-8\right)\times10^{-4}$ & $\left(0.06-1\right)\times10^{-3}$ \\
 $\mathrm{initial}~{e}$ & $ 60.2 $  & $ 60.2 $ & $ 40.4 $ & $0.02-0.2$ & $0.002-0.02$ \\
 $\mathrm{final}~{e}$ & $ 60.2 $  & $ 60.2 $ & $ 40.0 $ & $0.03-0.2$ & $0.001-0.02$ \\ \hline
 $Q^2 = 1.2~\mathrm{GeV}^2$ &  &  &  &  &  \\ \hline
 $\nu_\mu$ & $10.4$  & $10.5$ & $5.0$ & $\left(0.2-2\right)\times10^{-3}$ & $\left(0.2-2\right)\times10^{-3}$  \\
 $\overline{\nu}_\mu$ &  $10.4$  & $10.5$ & $5.0$ & $\left(0.6-5\right)\times10^{-3}$ & $\left(0.6-6\right)\times10^{-3}$ \\
 $\mathrm{initial}~{e}$ & $ 66.6 $  & $ 66.3 $ & $ 40.4 $ & $0.007-0.06$ & $0.001-0.01$ \\
 $\mathrm{final}~{e}$ & $ 66.6 $  & $ 66.3 $ & $ 38.0 $ & $0.02-0.2$ & $\left(0.4-5\right)\times10^{-3}$\\ \hline
 \end{tabular}
 \caption{Radiative energy loss (its absolute value) in (anti)neutrino-nucleus and electron-nucleus scattering for the radiation of soft photons at all orders in the opacity expansion at leading order in $\alpha$($\Delta E^{1\gamma}_S$) and after the resummation of leading logarithms ($\Delta E^\gamma_S$), for the radiation of one collinear photon in vacuum within the cone of $\Delta \theta = 10^\circ$ size ($\Delta E^{1\gamma}_c$) and for the radiation of one soft ($\Delta E^{1}_S$) or collinear ($\Delta E^{1}_c$) photon at the first order of the opacity expansion. The incoming (anti)neutrino and electron beam energy is $2~\mathrm{GeV}$. Two numbers for the first order of the opacity expansion correspond to the $^2_1\mathrm{H}$ and $^{208}_{82}\mathrm{Pb}$ nuclei, respectively, with the typical uncertainty due to missing QED higher-order contributions beyond the quoted significant digits.} \label{tab:energy_losses}
\end{table}
\end{center}

\section{Conclusions and Outlook} \label{sec:conclusions}

In this paper, we evaluated the QED nuclear medium-induced soft and collinear radiation of photons in (anti)neutrino-nucleus and electron-nucleus interactions at leading $\mathrm{SCET}_\mathrm{G}$ power and estimated the radiative energy loss of electrically charged leptons inside the nuclear medium. We presented expressions for all orders in the opacity expansion for both soft and collinear radiation and resummed the nuclear medium-induced radiation up to all orders of the opacity expansion, also including (for the soft radiation) double-logarithmic and soft-photon energy cutoff suppressions up to all orders of the electromagnetic coupling constant. The medium-induced soft and/or collinear radiation shifts the normalization of the broadening distributions of the charged leptons traveling inside the nucleus with double-logarithmically suppressed corrections. At each order of the opacity expansion, the relative contribution of soft or collinear radiation to the total bremsstrahlung coincides with the QED nuclear medium correction to the radiation-free process, which results in the vanishing QED nuclear medium-induced radiation after the resummation of all orders of the opacity expansion, at leading $\mathrm{SCET}_\mathrm{G}$ power. At the first order of the opacity expansion, we found that the radiative energy losses of charged leptons inside nuclear medium at GeV energies are below the typical experimental resolution. Our calculation provides the first-ever estimates of soft and collinear bremsstrahlung, as well as radiative energy losses of charged leptons inside the nucleus that are induced by the Glauber-photon interactions with the QED nuclear medium.

\section*{Acknowledgments}

OT acknowledges helpful discussions with Adi Ashkenazi. This work is supported by the U.S. Department of Energy through the Los Alamos National Laboratory. Los Alamos National Laboratory is operated by Triad National Security, LLC, for the National Nuclear Security Administration of the U.S. Department of Energy (Contract No. 89233218CNA000001). This research is funded by LANL’s Laboratory Directed Research and Development (LDRD/PRD) program under Project No. 20210968PRD4 and No. 20240127ER. FeynCalc~\cite{Mertig:1990an,Shtabovenko:2016sxi}, LoopTools~\cite{Hahn:1998yk}, Mathematica~\cite{Mathematica}, and DataGraph~\cite{JSSv047s02} were used in this work.

\appendix

\section{From the amplitude to cross-section corrections} \label{app:derivation}

In evaluating the squared matrix element for the bremsstrahlung with multiple exchange of Glauber photons between the charged lepton line and nuclear medium, there are contributions with the same space-time point of the Glauber interaction in direct and conjugated amplitudes, which we call ``contracted," and contributions  with photons at different space-time points in direct and conjugated amplitudes, which we call ``noncontracted". Each noncontracted contribution enters either with the factor $v\left(0 \right)$ for the exchange of just one Glauber photon or with a closed loop of $n$ Glauber exchanges in one space-time point, which does not change the kinematics of the hard process. Note that diagrams with the radiation of photons in between the Glauber exchanges in the single space-time point vanish. We verified explicitly that each noncontracted contribution reduces the number of Glauber momenta and coordinates in the function $\Gamma^{(n)}$ by one to $\Gamma^{(n-1)}$ and makes this function depend on the coordinates and Glauber momenta of contracted points. The one-Glauber exchange function $\Gamma^{(1)}$ reduces to $1$.

Exploiting this property, we further verified explicitly at the first three orders in the opacity expansion that all contributions with $v(0)$ and $v(0)^2$ vanish in the same groups of diagrams as was derived for the radiation-free process in Ref.~\cite{Tomalak:2023kwl}, with an extension by all possible insertions of one real photon. At each order of the opacity expansion, the same diagrams as  derived for the radiation-free process~\cite{Tomalak:2023kwl} contribute to the cross-section modification after adding diagrams with the coupling of one real photon to all sites before and after the exchange of each Glauber photon.

\section{Collinear radiation in the limit $x\to 1$} \label{app:limit_large_x}

In the limit of large relative photon energy $x \to 1$, for the functions $\Gamma^{(n)}$ in Eqs.~(\ref{eq:opacity_first_order_collinear})-(\ref{eq:opacity_n_order_collinear}) we obtain
\begin{align}
\frac{|\Gamma^{(1)} |^2}{{A}^2_{\perp,1}}  &= \left( \frac{\vec{A}_{\perp,1}}{{A}^2_{\perp,1}} - \frac{\vec{A}_{\perp,2}}{A^2_{\perp,2}}  \right)^2 + \frac{1}{A^2_{\perp,2}},  \\
\frac{|\Gamma^{(2)} |^2}{{A}^2_{\perp,1}}  &=  \left( \frac{\vec{A}_{\perp,1}}{{A}^2_{\perp,1}} - \frac{\vec{A}_{\perp,2}}{A^2_{\perp,2}}  \right)^2 + \left( \frac{\vec{A}_{\perp,2}}{A^2_{\perp,2}} - \frac{\vec{A}_{\perp,3}}{A^2_{\perp,3}}  \right)^2 + \frac{1}{A^2_{\perp,3}}, \\
\frac{|\Gamma^{(n)} |^2}{{A}^2_{\perp,1}} &= \left( \frac{\vec{A}_{\perp,1}}{{A}^2_{\perp,1}} - \frac{\vec{A}_{\perp,2}}{A^2_{\perp,2}}  \right)^2 + ... + \left( \frac{\vec{A}_{\perp,n}}{A^2_{\perp,n}} - \frac{\vec{A}_{\perp,n+1}}{A^2_{\perp,n+1}}  \right)^2 + \frac{1}{A^2_{\perp,n+1}}.
\end{align}
Assuming $p^\prime_\perp \gg q_{i,\perp}$, the expression for any order of the opacity expansion at leading $\mathrm{SCET}_\mathrm{G}$ powers can be written as
\begin{align}
\frac{|\Gamma^{(n)} |^2}{{A}^2_{\perp,1}} &= \frac{1}{{A}^2_{\perp,1}} \left( 1 + \sum \limits_{i=1}^n \left( \frac{  q_{i,\perp} }{p^\prime_\perp} \right)^2\cos^2 \phi_{i,\perp} + ... \right), \label{eq:opacity_large_x_all_orders_collinear}
\end{align}
where $\phi_{i,\perp}$ is the angle between $\vec{p}^\prime_\perp$ and $\vec{q}_{i,\perp}$. Considering the leading term in this expansion, we recover the relative radiation-free QED nuclear medium corrections at each order of the opacity expansion.

The limit of large photon energy is valid only very close to the photon end point when $1 - x \ll \frac{k^2_{\gamma,\perp} R_\mathrm{rms}}{\left(p^\prime + k_\gamma \right)^+}$. In the limit $x \to 1$, power corrections in Eq.~(\ref{eq:opacity_large_x_all_orders_collinear}) can introduce an additional collinear singularity resulting in a larger relative QED nuclear medium contribution at $x\to1$. For the radiation from the final-state charged lepton, such enhancement can be observed only at forward kinematics for a relatively small phase-space region due to the kinematic constraint $1-x \ge \frac{m_\ell}{E_\mathrm{beam}} \left[ 1 + \mathrm{O} \left( \frac{m^2_\ell}{M E_\mathrm{beam}} \right) \right]$. For the radiation from the initial-state charged lepton, the enhancement can be observed at arbitrary scattering angles but also in restricted phase-space region $1-x \ge \frac{m_\ell}{E_\mathrm{beam}} \left[ 1 + \mathrm{O} \left( \frac{m^2_\ell}{E_\mathrm{beam}^2} \right) \right]$.

\section{Collinear radiation in the region of intermediate $x$} \label{app:limit_intermediate_x}

In the limit of small arguments of the phases in Eqs.~(\ref{eq:opacity_first_order_collinear})-(\ref{eq:opacity_n_order_collinear}), corresponding to large photon formation times relative to the nuclear size, we expand in series of this small argument and obtain
\begin{align} 
|\Gamma^{(1)} |^2  &= 1 + \left( \vec{A}_{\perp,1} -\vec{A}_{\perp,2}  \right) \cdot \vec{A}_{\perp,1}  \left(\frac{{A}_{\perp,2} \left( z^1 - z^0 \right)}{ x\left(p^\prime \right)^+} \right)^2, \label{eq:opacity_intermediate_x_first_order_collinear} \\
|\Gamma^{(2)} |^2  &= 1 + \left( \vec{A}_{\perp,1} - \vec{A}_{\perp,3}  \right) \cdot \vec{A}_{\perp,1}  \left(\frac{{A}_{\perp,3} \left( z^1 - z^0 \right)}{x\left(p^\prime \right)^+} \right)^2 + \left( \vec{A}_{\perp,1} - \vec{A}_{\perp,2}  \right) \cdot \vec{A}_{\perp,1}  \left(\frac{{A}_{\perp,2} \left( z^2 - z^1 \right)}{x \left(p^\prime \right)^+} \right)^2 \nonumber \\
&+ 2 \left( {A}^2_{\perp,1} \vec{A}_{\perp,2} - A^2_{\perp,2} \vec{A}_{\perp,1}  \right)  \cdot \vec{A}_{\perp,3} \frac{\left( z^2 - z^1 \right)  \left( z^1 - z^0 \right) }{\left( x\left(p^\prime \right)^+ \right)^2},  \label{eq:opacity_intermediate_x_second_order_collinear} \\
|\Gamma^{(n)} |^2 &= 1 + \sum \limits_{i = 1}^{n+1}  \left( \vec{A}_{\perp,1} -\vec{A}_{\perp,i}  \right) \cdot \vec{A}_{\perp,1}  \left( \frac{ {A}_{\perp,i} \left( z^{n + 2 - i} - z^{n + 1 - i} \right)}{ x\left(p^\prime \right)^+} \right)^2 \nonumber \\
&+ 2 \sum \limits_{i = 2}^{n+1} \sum \limits_{j = 1}^{i-1} \left( {A}^2_{\perp,1}  \vec{A}_{\perp, j} - {A}^2_{\perp, j} \vec{A}_{\perp, 1}  \right)  \cdot \vec{A}_{\perp,i} \frac{\left( z^{n + 2 - i} - z^{n + 1 - i} \right) \left( z^{n + 2 - j} - z^{n + 1 - j}   \right)}{\left( x\left(p^\prime \right)^+ \right)^2}. \label{eq:opacity_intermediate_x_n_order_collinear}
\end{align}
Considering the first term in the expansion of the $\Gamma^{(n)}$ functions, i.e., $\Gamma^{(n)} \approx 1$, we recover the relative radiation-free QED nuclear medium corrections at each order of the opacity expansion. For this contribution, the relation between radiative QED nuclear medium effects and bremsstrahlung in vacuum for very large formation times is the same as the relation in the soft-photon limit $x \to 0$ and is the same as the relation for radiation-free process. Noticing that the deviations of the $\Gamma^{(n)}$ functions from unity have smaller powers of $A_{\perp,1}$ in the denominator, we do not expect an additional collinear enhancement from power-suppressed terms in Eqs.~(\ref{eq:opacity_intermediate_x_first_order_collinear})-(\ref{eq:opacity_intermediate_x_n_order_collinear}). Moreover, the power-suppressed terms do not generate logarithms of the ratio between atomic and nuclear scales and are expected to be small in agreement with our numerical estimates.

\newpage

\newpage
\bibliography{paper}{}

\end{document}